%% file: v0.tex
\documentclass[12pt, a4]{article}
\usepackage{graphicx,color}
\usepackage{graphics}
\usepackage{amssymb}
\usepackage{epstopdf}
\usepackage{amsmath}
\usepackage{overpic}
\usepackage{lineno}
\usepackage{verbatim}
\usepackage[dvipsnames]{xcolor}
\usepackage{graphicx,subcaption,color}

\RequirePackage{xspace}

\title{Brief Article}
\author{E.K.}
\newcommand{\be}{\begin{equation}}
\newcommand{\ee}{\end{equation}}
\newcommand{\bea}{\begin{eqnarray}}
\newcommand{\eea}{\end{eqnarray}}

\begin{document}
\def\red#1{\textcolor{red}{#1}}

\def\Bbar{\overline{B}^0}

\def\Dbar{{D}^{0}}
\def\Dstbar{{D}^{*0}}
\def\DDstbar{{D}^{(*)0}}
\def\Kbar{\overline{K}^0}
\def\Kstbar{\overline{K}^{*0}}
\def\KKstbar{\overline{K}^{(*)0}}
\def\piz{\pi^0}
\def\rhoz{\rho^{0}}

\def\Bm{{B}^-}
\def\Dm{{D}^{-}}
\def\Dstm{{D}^{*-}}
\def\DDstm{{D}^{(*)-}}
\def\DDsts{{D}^{(*)}_s}
\def\Km{{K}^-}
\def\Kstm{{K}^{*-}}
\def\KKstm{{K}^{(*)-}}
\def\pim{{\pi}^-}
\def\rhom{{\rho}^-}

\def\axion{a_{xion}}

\def\alps{``$a$''\ }
\def\bratio{{\rm Br} }

\def\GeV{{\rm\ GeV}}
\def\Slash#1{{#1}\mkern-7.5mu/}

\def\CP{\ensuremath{C\!P}\xspace}

\bibliographystyle{unsrt}
\renewcommand{\thefootnote}{\fnsymbol{footnote}}
\rightline{\today}
\vspace{1.5cm} 
\begin{center}
{\LARGE \bf  Belle II observation prospects for axion-like particle production from $B$ meson annihilation decay } 
\end{center}
\vspace{.3cm}

\begin{center} 
\sc{ \large $^{a,d}$Y.~Zhang,  $^{b,c}$A.~Ishikawa, $^d$E.~Kou, $^e$D.~Marcantonio, and $^e$P.~Urquijo} \\
\vspace{1cm}
\sl{\small $^a$Key Laboratory of Nuclear Physics and Ion-beam Application (MOE), Fudan University, 200433 Shanghai, China }\\
\sl{$^b$ High Energy Accelerator Research Organization (KEK), 305-0801 Tsukuba, Japan}\\
\sl{\small $^c$ SOKENDAI (The Graduate University for Advanced Studies), Hayama 240-0193, Japan}\\
\sl{\small $^d$ Universit\'{e} Paris-Saclay, CNRS/IN2P3, IJCLab, 91405 Orsay, France }\\
\sl{\small $^e$ School of Physics, The University of Melbourne, Victoria 3010, Australia}\\
\end{center}

\vspace{1cm}
\begin{abstract}
We investigate a new production mechanism of axion-like particle (ALP) from $B$ meson annihilation decays and its observation potential at the Belle and Belle II experiments. 
This mechanism allows for the production of ALP from $B$ meson decays in association with a large variety of mesons. 
In this article, we first estimate the branching ratios of such processes with a perturbative QCD method. 
Focussing on the most promising $B\to h a'$ ($h=K^\pm, \pi^\pm, D^0$ and $D_s$) channels, we perform sensitivity studies for $a'$ decaying invisibly or into diphoton with Belle and Belle II experiments. 
\end{abstract}

\newpage
\section{Introduction}
\input{intro}

\section{Annihilation diagram computation in pQCD}\label{section:2}

\subsection{ALP production from annihilation diagrams}
\input{diagram-annihilation}

\subsection{Computation of the ALPs production from the annihilation diagram in pQCD method }
\input{pQCD-0}

\section{Belle (II) sensitivity study}
\input{Belle}

\subsection{Invisible decay of ALP}
\input{invisible-Belle}

\subsection{Di-photon decay of ALP}
\input{diphoton_mode}

\subsection{Observation potentials at Belle and Belle II}
\input{Belle-BelleII}
\section{Conclusions}
In this article, we investigated ALP production from the weak annihilation transition in $B$ meson decays. First, we have shown that a large variations of the  final states is possible from this new mechanism, i.e. $B\to ha'$, where $h=D^{0(*)}, K^{\pm (*)}, K^{0, (*)}, D_s^{(*)},$ $\pi^\pm, \rho^\pm, \pi^0, \rho^0, \phi,$ $J/\psi, \eta_c, \cdots$.  

We performed a pQCD based theoretical computation to predict the branching ratios of these processes. It is found that assuming that the ALP-fermion coupling to be $g_i=1$, the branching ratios are roughly $10^{-6}$  for the models where ALP couples to the light quarks (i.e. up, down, strange), $10^{-7}$ for the charm quark, and $10^{-9}$ for the bottom quark. The suppression for the bottom quark is the well known $1/m_b$ effect. Based on this computation, we conclude that the four channels with  large branching ratios,  $B\to ha'$ with $h=D^{0}, K^{\pm}, D_s,  \pi^\pm$, and  we next performed sensitivity studies for these channels with Belle (II).  

We focused on the ALP decays into invisible or di-photon final states, which are the strength of the Belle (II) experiment. In order to obtain the achievable upper limit on the branching ratios, 
we performed detailed sensitivity study using Belle MC. 
In both final states, the upper limit of the branching ratios for the $K^\pm$ and $\pi^\pm$ channels are an order of magnitude better than the ones for the $D^0$ and $D_s$ channels. 
This is partially because the latter further decays weakly and only a part of their decay channels can be considered as the rest of the decay channels are higher multiplicity channels and difficult to analyse (i.e. much higher background). 
The $K^\pm$ and $\pi^\pm$ channels are sensitive to the  models where ALP couples to up, down or strange quarks. Thus, 
we found that those models could be observed at the Belle (II) experiment. 
On the other hand, although our theoretical prediction shows that the $D^0$ and $D_s$ final states are the most promising channels for the models with charm or bottom quarks, the predicted branching ratio is too small to be observed at Belle (II). However, these channels are still interesting to investigate for the models where the coupling of quarks and ALP are induced by the derivative coupling, for which the coupling is proportional to the quark masses and is enhanced for heavy quarks.

\section*{Acknowledgement}
We would like to acknowledge J. Zupan for his collaboration at the early stage of the project and his careful reading of the manuscript. The work of A.~I. and E.~K.  is supported by TYL-FJPPL. A.~I. is supported by JSPS KAKENHI Grant Number 22H00144. Y.~Zhang is supported by China Scholarship Council (CSC) under the State Scholarship Fund No.~201906100100. 
\clearpage

\newpage
\input{pQCD}


\end{document}

%% file: intro.tex
It has been a long time since the axion particle has been introduced by Peccei and Quinn as a solution to the strong \CP problem~\cite{quinn:1977}, which questions why the \CP violation allowed in the Standard Model (SM) is extremely suppressed in  nature as seen from the non-observation of the neutron electric dipole moment. 
When we consider this suppression as a result of the underlying global $U(1)$ symmetry, the axion particle emerges as the Nambu-Goldstone boson of the spontaneous breaking of this $U(1)$ symmetry. 
In this article, we investigate the so-called axion-like particle (ALP), a generic name for pseudoscalar particles at a GeV scale, much heavier than the axion in the original model~\cite{Georgi:1986df}-\cite{Gavela:2019}. 
The purpose of this article is to investigate a novel ALP production mechanism, $B$ meson annihilation decay, and its search potential at the Belle (II) experiment. 

In refs~\cite{Torben:2022}-\cite{Motoi:2020}, it has already been shown that the Belle (II) experiment can provide a unique probe for ALP searches. ALP searches are being intensively pursued, especially in the early physics program of Belle II, the specialised study before the SuperKEKB accelerator reaches luminosities high enough to perform its flavour physics programs, e.g. with $B, D$ mesons or $\tau$ leptons. For example, the $e^+e^- \to 3 \gamma$ process where ALP could be produced through a $a' \gamma\gamma$ coupling, a specialised trigger was set up~\cite{Belle2:2020}. 
Taking advantage of its clean environment of the $e^+e^-$ collider, Belle (II) can perform very unique searches for ALP which decay into $\gamma\gamma$ or nothing, i.e. invisible decay, even with existing data sets. Thus, we focus on these decay channels in this article. 

In the following, we investigate the novel production mechanism of ALP from the $B$ meson annihilation process. The annihilation process occurs as the constituent quarks of $B$ mesons (anti-bottom and up quark for $B^+$ meson and anti-bottom and down quark for $B^0$ meson) annihilate via the four-fermion interaction and produce hadrons in the final state. 
On one hand, the observations of pure annihilation processes, $B^0\to K^+K^-$ ($5.2\sigma$ significance)~\cite{BtoKK:2012,BtoKK:2017} or $B_s\to \pi^+\pi^-$ ($7.0\sigma$ significance)~\cite{Bstopipi:2012}, prove the existence of such process. 
On the other hand, annihilation processes have played important roles in $B$ physics, such as the isospin violation of the $B\to \rho \gamma$ process~\cite{Beyer:2001}-\cite{Cai-Dian:2005} or the strong \CP phase generated by the annihilation being the important source of the direct \CP violation in the charmless $B$ decays~\cite{Keum:2001,Bauer:2004}. 
In this paper, we investigate the phenomenology of the annihilation process in ALP search at Belle (II) experiment. The mechanism of producing ALP we consider here is that the ALP is emitted from one of the quark lines of the four-fermion interaction, as shown in Fig~\ref{fig:1}. 
ALP production from $B$ mesons has been discussed in many articles but they mainly consider ALP production from the quarks in the loop (typically, top quark) or from the $W$ boson. 
Thus, the final states are limited to $B\to K^{(*)} a'$ or $B\to \pi (\rho) a'$. The advantage of the annihilation mechanism is that the ALP can be produced associated with many other hadrons, such as $D, D^{*}, D_s\cdot$ as well as charmonium, $J/\psi, \eta_c \cdots$, which opens more channels to explore at Belle (II). 
Furthermore, it can occur from tree level process, whose branching ratio, in general, is larger than the loop level penguin process. 

The annihilation diagram is calculable using either the so-called pQCD method~\cite{Keum:2004} or the QCD factorisation method~\cite{Beneke:2004}. 
The computation is very close to the annihilation contributions to  the radiative $B\to \rho \gamma$ decay~\cite{Stefan:2002,Cai-Dian:2005}. 
In these approaches, the initial and the final state meson distribution amplitudes are convoluted with a hard kernel, which, in this case,  results in the $a'$ emission. 
The hard kernel includes the propagators of the quarks, which are between the $a'$ emission and the four-fermion interaction. 
Although the annihilation diagram is a $1/m_b$ contribution and is generally suppressed, when there is a light quark propagator (i.e. ALP emission from light quark), it is enhanced by a factor of $1/\Lambda_{\rm QCD}$. 
In addition, the ALP-quark coupling is   proportional to the quark mass in the models where the ALP-fermion coupling is induced by derivative and this effect could compensate the $1/m_b$ suppression factor for the heavy quark cases. 
Thus, we will consider emission of ALP from all possible quarks in $B$ decay, i.e. $u, d, s, c, b$ and perform the sensitivity study for the best observation channels.   

The article is organised as follows. In section 2, we describe our theoretical framework for ALP interactions, their couplings and the mass range we are interested in. In section 3, we review the computation of ALP production from the $B$ meson annihilation diagram using the pQCD method. In section 4, we present our sensitivity study at Belle (II) experiment and we conclude in section 5.

%% file: diagram-annihilation.tex
In this subsection, we show all the Feynman diagrams that can produce ALP from annihilation $B$ decays. 
We first categorise the final states by the corresponding CKM matrix elements as well as the colour-allowed tree, colour-suppressed tree or penguin topologies. 
For the purpose of this section, it is enough to consider only the dominant contributions. That is, $B\to Ka'$ and $B\to \pi a'$ have both penguin and tree contributions but they are, respectively, considered as the penguin and the tree processes. Similarly, the charmonium final state is considered as a tree process. 
The result, taking into account diagrams up to $\lambda^3$ order, is listed on the Table~\ref{tab:pQCD_cal}. 
Naively we expect larger branching ratios for the tree diagrams. 
Among them, we investigate those with Cabbibo allowed colour suppressed tree processes, and Cabbibo suppressed colour allowed tree processes in the following, i.e. $D^0, D^{*0}, D_s, D_s^{*}, \pi^+, \rho^+$ final states. In addition, we also include Cabbibo allowed penguin processes, i.e. $K^+, K^{*+}, K^0, K^{*0}$, as their experimental sensitivity is known to be quite high. 
The charmonium production is suppressed with respect to the two tree processes mentioned above but it can be interesting for future study. 
It should be mentioned that the pQCD method is not the most suitable for charmonium production and we would need to perform an independent theoretical investigation for these processes. 

\begin{figure}[htbp]
\begin{center}
\includegraphics[height=5.0cm]{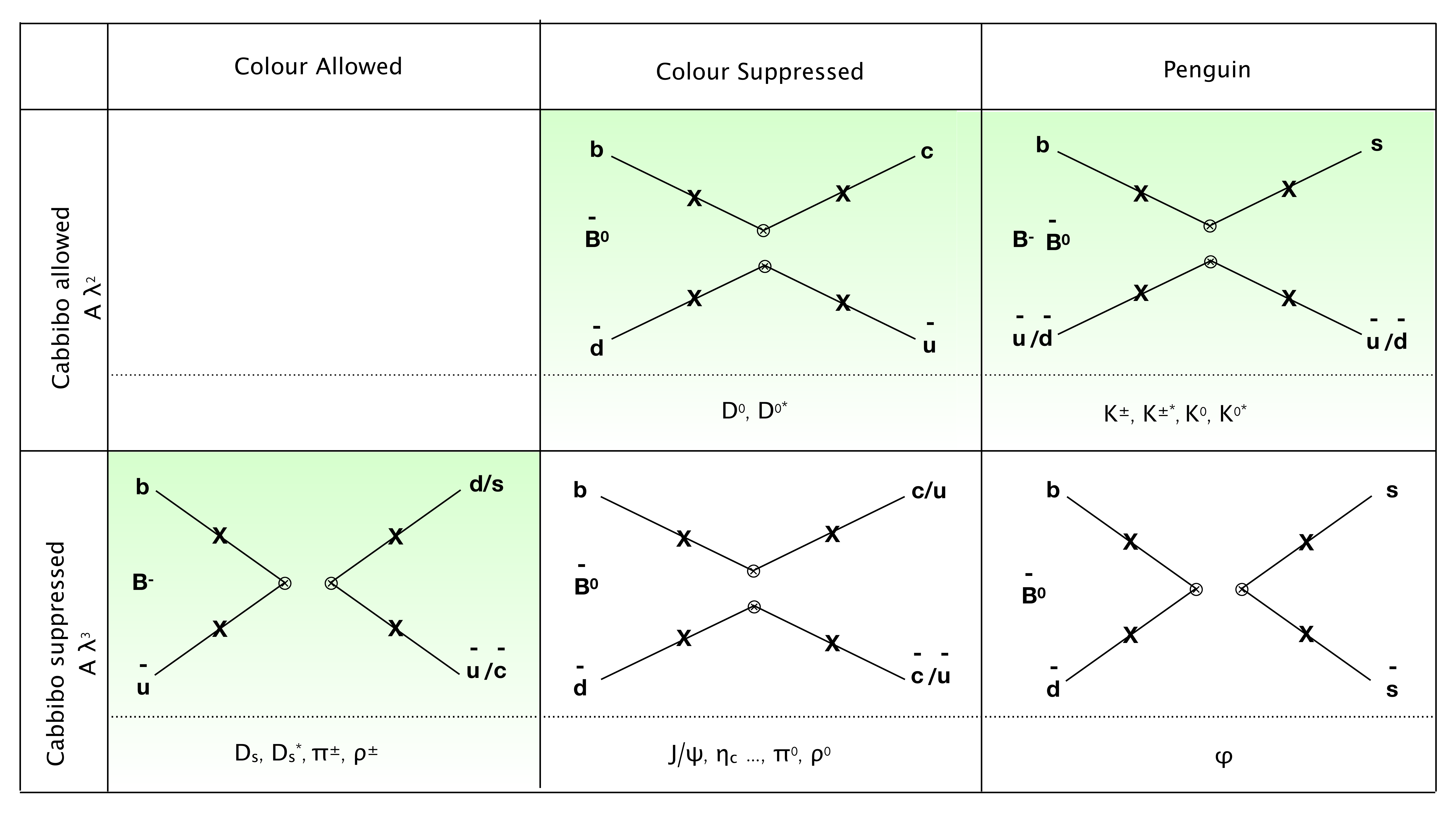}
\caption{ALP production from $B$ meson annihilation decays: ALP can be produced at any of the quark line (cross mark), depending on the ALP coupling to different type of quarks. In this figure, we categorise the annihilation diagrams by the Cabbibo factors as well as colour allowed/suppressed tree or penguin topologies.}
\label{fig:1}
\end{center}
\end{figure}

%% file: pQCD-0.tex
In this section, using the $\Bbar \to \Dbar a'$ process as an example, we demonstrate how to compute ALP emission from the annihilation diagram in the pQCD method. 

Let us start with the weak Hamiltonian:
\be
H^{\rm eff}=\frac{G_F}{\sqrt{2}} \sum_{i=1,2} V_{cb}^*V_{ud}C_i O_i^c  
\ee
where $G_F$ is the Fermi constant and the $V_{qq'}$ is the CKM matrix element. $C_i$ is Wilson coefficient. The operators $O_i^c$ are defined as 
\be
O_1^c=(\overline{u}_id_j)_{V-A}(\overline{c}_jb_i)_{V-A}, \quad O_2^c=(\overline{u}_id_i)_{V-A}(\overline{c}_jb_j)_{V-A}
\ee
where $(\overline{\psi}_i\psi^\prime_j)_{V-A}=\overline{\psi}_i\gamma^\mu(1-\gamma_5)\psi^\prime_j$ and $(i,j)$ are the colour indices. 
For annihilation decay, the initial $b$ quark and $d$ quark must be in the same current. Thus, we use the Fiertz transformed operators 
\be
O_1^{c{\rm F}}=(\overline{c}_ju_j)_{V-A}(\overline{d}_ib_i)_{V-A}, \quad O_2^{c{\rm F}}=(\overline{c}_ju_i)_{V-A}(\overline{d}_ib_j)_{V-A}.
\ee

Then, the amplitude is given as 
\be
A^{1(a,b)}=\frac{G_F}{\sqrt{2}}V_{cb}^*V_{ud}{a_1}\langle \Dbar a' | O_1^{c{\rm F}} | \Bbar \rangle^{(a,b)}, \quad 
\ee
where  $a_1=C_1+C_2/N_C$. The indices $a$ and $b$ indicate ALP emission in Fig~\ref{fig:1} from the initial $b, \overline{d}$ and the final $c, \overline{u}$ quarks, respectively. 

Using the definitions of decay constants, distribution functions and form factors given in Appendix A, as well as the axion coupling defined as 
\be
{\mathcal{L}}=g_q a'\overline{q}\gamma_5q; \quad q=u,d,s,c,b,  
\ee
we obtain the amplitudes for $\Bbar \to \Dbar a'$: 
 \bea
 A^{1a}(\Bbar\to \Dbar a') &=& \frac{G_F}{\sqrt{2}}V_{cb}^*V_{ud} (i f_D) {a_1}\left[f_1^{D^0} (p_D\cdot p_D)+f_2^{D^0} (p_D\cdot p_{a'} )\right], \\
 A^{1b}(\Bbar\to \Dbar a') &=& \frac{G_F}{\sqrt{2}}V_{cb}^*V_{ud}  (-i f_B){a_1} \left[f_3^{D^0} (p_B\cdot p_D)+f_4^{D^0} (p_B\cdot p_{a'} )\right], 
  \eea
 where the $Ba'$ form factors ($f_{1,2}$) and  $Da'$  form factors ($f_{3,4}$) are computed as a convolution of the distribution function of the $B$ and $D$ mesons and the hard kernel thst represents the $a'$ emission.\footnote{The $K_0$ and $H_0$ terms represent the Bessel and the Hankel functions, which appear by the Fournier transformation of the $k_\perp$ of the spectator to the impact parameter $b$, which leads to  $\int^{\infty}_{-\infty}dk^2_\perp \frac{1}{k_\perp^2+a^2}=\int_0^{\infty} d|b| |b| K_0(a|b|)$ and $\int^{\infty}_{-\infty}dk^2_\perp \frac{1}{k_\perp^2-a^2}=\int_0^{\infty} d|b| |b| \frac{i\pi}{2} H_0(a|b|)$.} The results yield 
\bea
 {a_1}f_1^{D^0}   &=& 
  \int_0^1 dx \int_0^\infty db  \bigg[ (ig_d)N_c a_1(t)\left(\frac{-i}{\sqrt{2N_c}}\Phi_B(t)\right)  \nonumber\\
  & & \times (-1) 4m_B (-x_1)|b_1|K_0(m_B\sqrt{x_1}, |b_1|)  \bigg]  \label{eq:8} \\
&+& \int_0^1 dx \int_0^\infty db  \bigg[ (ig_b)N_c a_1(t)\left(\frac{-i}{\sqrt{2N_c}}\Phi_B(t)\right)  \nonumber\\
& & \times (-1) 4m_B (1-x_1)|b_1|K_0(m_B\sqrt{1+x_1}, |b_1|)  \bigg] \nonumber\\
    {a_1}f_2^{D^0}   &=& 
  \int_0^1 dx \int_0^\infty db  \bigg[ (ig_d)N_c a_1(t)\left(\frac{-i}{\sqrt{2N_c}}\Phi_B(t)\right)  \nonumber \\
  & &  \times (-1) 4m_B |b_1|K_0(m_B\sqrt{x_1}, |b_1|)  \bigg] \label{eq:9}\\
   {a_1}f_3^{D^0}   &=& 
    \int_0^1 dx \int_0^\infty db   \bigg[ (ig_u)N_c a_1(t)\left(\frac{i}{\sqrt{2N_c}}\Phi_D(t)\right)  \nonumber\\
    & & \times (-1) 4m_D(-x_2)|b_2|\frac{i\pi}{2}H_0(m_B\sqrt{x_2}, |b_2|)  \bigg] \label{eq:10}\\
&+& \int_0^1 dx \int_0^\infty db  \bigg[ (ig_c)N_c a_1(t)\left(\frac{i}{\sqrt{2N_c}}\Phi_D(t)\right)  \nonumber \\
 & &  \times (-1) 4m_D(1-x_2)|b_2|\frac{i\pi}{2}H_0(m_B\sqrt{1-x_2}, |b_2|)  \bigg] \nonumber\\
   {a_1}f_4^{D^0}   &=& 
    \int_0^1 dx \int_0^\infty db  \bigg[ (ig_u)N_c a_1(t)\left(\frac{i}{\sqrt{2N_c}}\Phi_D(t)\right)  \nonumber\\
  & &  \times (-1) 4m_D(-)|b_2|\frac{i\pi}{2}H_0(m_B\sqrt{x_2}, |b_2|)  \bigg] \label{eq:11} \\
&+&  \int_0^1 dx \int_0^\infty d|b|  \bigg[ (ig_c)N_c a_1(t)\left(\frac{i}{\sqrt{2N_c}}\Phi_D(t)\right)  \nonumber \\
  & &  \times (-1) 4m_D|b_2|\frac{i\pi}{2}H_0(m_B\sqrt{1-x_2}, |b_2|)  \bigg] \nonumber
\eea   
where $x_{i}$ and $b_{i}$ are the parameters  describing the momentum of the light quarks insides of the $B$ and $D$ mesons ($i=1$ for $B$ and $i=2$ for $D$, find notations e.g. in~\cite{Cai-Dian:2005,Keum:2001}).
The integration of $x$ and $|b|$ must be taken in the range $\frac{\sqrt{2}}{m_B}\Lambda_{QCD}<x<1, 0<|b|<\frac{1}{\Lambda_{QCD}}$ after multiplying though the Wilson coefficient, 
\be
a_1(t)\quad {\rm with}\quad   t={\rm max}[\sqrt{X}m_B,1/b]
\ee
where $X$ represents the arguments of the Bessel functions, $K_0, H_0$, in Eqs. (\ref{eq:8}-\ref{eq:11}), i.e. $x_1, 1+x_1, x_2, 1-x_2$.
As is well known, the $a_1$ value depends strongly on the renormalisation scale, $t$, e.g. from $\sim -0.32$ to $\sim + 0.06$ for $t=1-5$ GeV. In pQCD, $t$ varies according the the momentum fraction carried by the light quark in the mesons. 

Equations (\ref{eq:8}-\ref{eq:11})  show that this process is sensitive to the $g_{u, d, c, b}$ couplings. As mentioned in the introduction, the dominant contribution comes from the $a'$ emission from the spectator, namely the $f_2^{D^0}$ term. This is because  the distribution function of the $B$ meson peaks at smaller value of $x_1$ while the Bessel function $K_0$ is suppressed at a higher value of $x_1$. 

The amplitudes for the Cabibbo allowed tree processes ($\DDstm, \DDsts$) are computed in a similar way, with Wilson coefficient $a_1(t)$, instead of $a_2(t)$, which varies more moderately from $\sim 1.2$ to $\sim 1.0$ for $t=1-5$ GeV. 
These processes are sensitive to the $g_{u, d, c, b}$ and $g_{u, s, c, b}$ couplings, respectively. 
The ($\pi^\pm$/$\rho^\pm$, $\KKstbar$, $\KKstm$) processes come from both penguin and tree diagrams and receive contributions from $a_2(t), a_4(t), a_6(t)$. The computation for $B \to K a'$ is given in Appendix B as an example.  They are sensitive to the $g_{u, d, b}$, $g_{u, s, b}$ and $g_{d, s, b}$ couplings, respectively. 

\subsection{Which final states are to be searched at Belle (II)?}

\begin{table}[]
    \centering
    \caption{The pQCD computation result of $Br(B\to h a')g_q^2$ of the $B\to ha' $ processes (where $h=D^0, D^{*0}, D_s, D_s^{*}, K^\pm, K^{*\pm}, K^0, K^{*0}, \pi^\pm, \rho^\pm$) in the unit of $\times 10^{-6}$. }
    \begin{tabular}{|l|ccccc|}
    \hline
         & up & down & strange & charm & bottom   \\ \hline
  $D^{0}$& 0.270 & 3.700 & 0.000 & 0.062 & 0.004 \\
  $D^{*0}$& 0.155 & 3.462 & 0.000 & 0.011 & 0.000 \\  
  $D_{s}$& 3.573 & 0.000 & 0.305 & 0.129 & 0.002\\
  $D^{*}_{s}$& 3.403 & 0.000 & 0.066 & 0.025 & 0.000\\
  $K^{\pm}$& 2.860 & 0.000 & 1.659 & 0.000 & 0.001\\
  $K^{*\pm}$& 0.765 & 0.000 & 1.370 & 0.000 & 0.000\\
  $K^{0}$& 0.000 & 2.220 & 1.510 & 0.000 & 0.0012\\
  $K^{*0}$& 0.000 & 1.090 & 1.590 & 0.000 & 0.000\\
  $\pi^{\pm}$& 2.080 & 0.320 & 0.000 & 0.000 & 0.000025\\
  $\rho^{\pm}$& 2.810 & 0.175 & 0.000 & 0.000 & 0.000062\\
  \hline
    \end{tabular}
    \label{tab:pQCD_cal}
\end{table}

%
%
%
%

Now let us see which processes are more sensitive to each $g_i$ coupling. The obtained results are presented in Table~\ref{tab:pQCD_cal}. 

Let us consider different $g_q\neq 0 $ scenarios and look into each cases to decide which channels are most promising for the Belle II experiment. 

\begin{description}
\item{$g_u\neq 0$:} 
In this case, the $D_s, D_s^*, K^\pm, \rho^\pm, \pi^\pm $ channels obtain equally large contributions. The channel with a $D_s^*$ is more challenging to reconstruct as it decays via $D_s^{*}\to D_s \gamma$ with a relatively low energy photon (over 90\% branching ratio). The $\rho^\pm$ decays into two pions, with one of them neutral and its experimental measurement suffers from much more background due to the broad width.
Thus, the best channels would be $B\to (D_s, K^\pm, \pi^\pm) a'$. 
\item{$g_d\neq 0$:} 
In this case, $D^0$ and $D^{*0}$ obtain large contributions. However, the channel with a $D^{*0}$ decaying to $D^0\pi^0$ or $D^0\gamma$ makes it more challenging than $D^0$ to be observed. So the best channel would be the $B\to D^0 a'$ for this case. 
\item{ $g_s\neq 0$:}
In this case, $K^+, K^{*+}, K^0, K^{*0}$ obtain the largest contributions. Experimentally the best option would be the $B\to K^\pm a'$ channel. 
\item{$g_c\neq 0$:}
In this case, the best channel seems to be $B\to D_s a'$. 
\item{$g_b\neq 0$:}
In this case, the best channel is $B\to D^0 a'$, where $D$ decays to $K^{-}\pi^{+}$, $K^{-}\pi^{+}\pi^{0}$ or $K^{-}\pi^{+}\pi^{+}\pi^{-}$.  
\end{description}


%% file: Belle.tex
In this section, we perform a Monte Carlo (MC) study to estimate the sensitivity of the ALP search with the Belle and Belle II experiments at the KEKB/superKEKb $e^+e^-$ colliders~\cite{Belle_detector}-\cite{Belle2det}. $e^+e^-$ colliders have an advantage over hadron machines, for missing energy and photon final states. Thus, we concentrate on the ALP decays, $a'\to$ invisible and $a' \to \gamma\gamma$ in this section. In the following, we perform a sensitivity study based on Belle MC, which can be extrapolated to Belle II. 

The Belle detector~\cite{Belle_detector} located at the interaction point (IP) of the KEKB collider, which operated at a centre-of-mass energy of 10.58 GeV. From 1999 to 2010, the Belle experiment collected an integrated luminosity of about 711 fb$^{-1}$. The detector~\cite{Belle_detector} consisted of concentric cylindrical subdetectors; a silicon vertex detector, a central drift chamber, particle ID detectors, an electromagnetic calorimeter (ECL), and a $K_L^0 -$ muon detector. In this work, the Belle detector geometry and response is simulated with \verb|Geant3|~\cite{GSIM}.

The decay channel of $B \rightarrow h a' $, where $h$ is $\pi,~K,~D^{0},~D_{s}$, are studied. The MC samples of $B^\pm \rightarrow \pi^\pm a^\prime,\ K^\pm a^\prime,\ D_{s}^\pm a^\prime$ and $\Bbar \to \Dbar a'$ are generated using \verb|EvtGen|~\cite{EVTGEN}. The $a^\prime$ is forced to promptly decay to neutrinos (to mimic a generic final state), or to a di-photon final state. The signal samples contain 1 million events for 33 different ALP mass hypotheses for $B \to D,~D_s$ and 45 mass hypotheses for $B \to \pi,~K$ processes due to the larger allowable mass range.

The background samples used in this work are MC samples produced by the Belle collaboration. They contain $e^+e^- \rightarrow B^+B^-,\ B^0\bar{B}^0$ (both at 10$\times$ the Belle integrated luminosity), and  $q\bar{q}$ (at 6$\times$ the Belle integrated luminosity) where $q = u,\ d,\ s,$ or $c$. There is also a sample with ``rare'' $B$ decays produced with $25$ times of the Belle integrated luminosity. This sample contains $B$ decays that despite having small or poorly measured branching fractions, could be important background to consider, e.g. $B^+ \rightarrow \pi^+ K^0$.

%% file: invisible-Belle.tex
Events are reconstructed by first converting Belle MC to a Belle II software (\verb|basf2|)~\cite{basf2} readable format using \verb|b2bii|~\cite{b2bii}. 
The signal-side $B$-meson ($B_{\rm{sig}}$) is reconstructed by reconstructing the signal-side hadron ($\pi$, $K$, $D_{s}$ or $D^0$). 
Charged particles (pions and kaons) are required to have their point of closest approach to the interaction point (IP) less than 2 cm and 3 cm in the directions transverse and parallel to the beam direction, respectively, as well as meet some particle identification requirements and have lab-frame momentum greater than 200 MeV/$c$. 
The $B \to D_s$ and $D$ channels use further selection on the invariant masses of combined tracks to find $D$ meson candidates, and a vertex fit is done on the $D$ candidate. 
The other $B$ meson ($B_{\rm{tag}}$) in the event is reconstructed with the Full Event Interpretation algorithm (FEI), which uses over 10, 000 channels 
We refer to the remaining tracks and calorimeter clusters not used in the reconstruction as the Rest of Event (ROE).

The $B_{\rm{tag}}$ is required to have a beam-energy-constrained mass of more than 5.27 GeV/$c^2$ and FEI must return a signal probability of greater than 0.005. 
The total unused ECL energy must be less than 0.4 GeV. Two multivariate classifiers are trained to separate the signal process from $q\bar{q}$ and non-signal $B$ events respectively. These are trained on variables that describe the shape of the events.


The search method looks for resonances in the signal-side hadron's $B_{\rm{sig}}$-frame momentum, which should peak due to two-body kinematics. The signal-side $B$ frame momentum is calculated by subtracting the four vectors of tag-side and ROE from that of the $e^+e^-$ beam.

The signal peak is fitted for each generated mass hypothesis using a convolution of two Gaussians. The free parameters of this probability density function (PDF) are then parameterised as a function of the mean of the Gaussians in order to have a well-defined signal PDF anywhere in the search range. The background distribution is parameterised using a kernel density estimator \cite{KDE:2001}. 

\begin{figure}[thb]
    \centering
    \includegraphics[width=10cm]{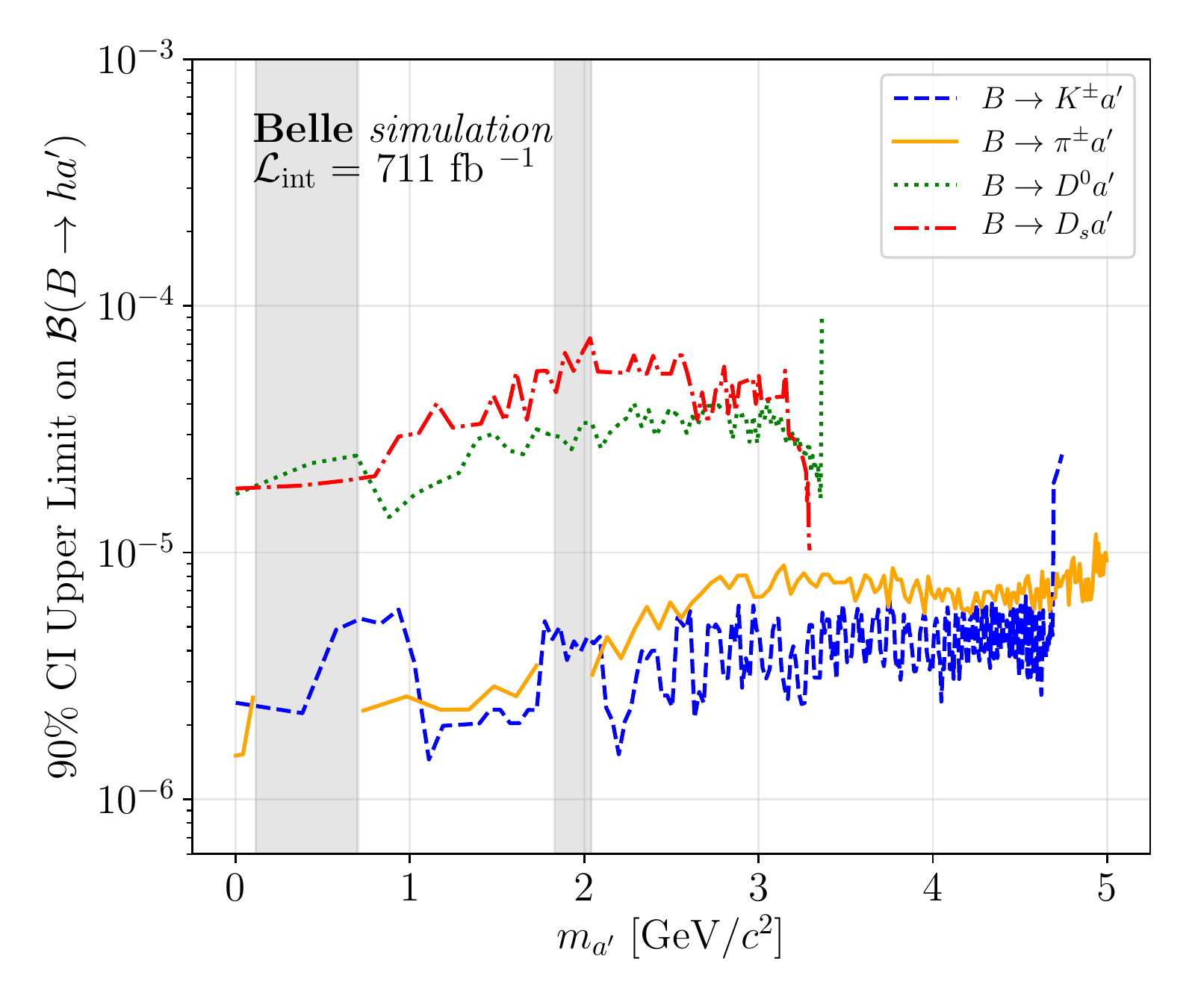}
    \caption{Expected upper limit for $B \rightarrow h  (a' \rightarrow \rm invisible)$ at Belle experiment, where $h$ is $\pi,~K,~D^{0},~D_{s}$. The filled grep bands are the veto ranges of $B \to K^{0} \pi$ and $B \to D^{(*)0} \pi$, respectively.} 
    \label{fig:sens_plot}
\end{figure}

A combined fit is performed over background-only MC, with only the yield of the signal and background PDFs allowed to float, and a 90\% confidence interval on the number of signal events yielded by the fit is calculated. We use this upper limit to estimate the branching fraction sensitivity of the analysis. Our result is shown in Fig.~\ref{fig:sens_plot}. The largest discrepancy in the sensitivity between MC and real data is driven by the difference in the FEI performance between MC and data. The efficiency of the FEI can be 30\% less on data compared to MC. The sensitivity limits shown here include a correction factor for the FEI efficiency derived from an analysis of $B^\pm \rightarrow K^\pm J/\psi$, $J/\psi \rightarrow \ell^+\ell^-$ decays where $\ell \in e,\mu$.


    
%% file: diphoton_mode.tex
In the $a' \to \gamma \gamma$ study, signal-side is fully reconstructed by combining a hadron $h =~\pi,~K,$ $D^{0},$ $D_{s}$ and a combination of two photons.
All charged particles are required to have distance from the $e^{+}e^{-}$ IP smaller than $0.2~\rm cm$ in the plane transverse to the beams and smaller than $1.0~\rm cm$ along the beam direction. The binary ratio
$\mathcal{L}(K/\pi) \equiv \frac{\mathcal{L}(K)}{\mathcal{L}(K) + \mathcal{L}(\pi)}$,
is used to identify the species of charged particles, where $\mathcal{L}(h)$
is the likelihood for a kaon or pion to produce the signals observed
in the detectors.  Charged particles with $\mathcal{L}(K/\pi) > 0.6$
are identified as kaons and those with $\mathcal{L}(\pi/K) > 0.6$ as
pions. The photons need to meet the criteria of energy $E>0.05~\rm{GeV}$.
The $\pi^{0}$ and ALP candidates are formed from the combination of two photons, the former within the invariant mass range of [0.125, 0.140] $\rm{GeV}$$/c^{2}$, corresponding to $\pm 2.5 \sigma $ away fron the nominal $\pi^{0}$ mass. 
To suppress peaking backgrounds originating from $\pi^{0}$ and $\eta$ in the mass spectrum of ALP candidates, we veto the di-photon mass ranges of [0.120, 0.160] $\rm GeV$/$c^{2}$ and [0.520, 0.590] $\rm GeV$/$c^{2}$.
An additional requirement on the $\pi^{0}$ momentum to be larger than $0.5~\rm{GeV}$$/c$ is implemented.
The $D$ mesons are reconstructed from the final states $K^{\mp}\pi^{\pm}$, $K^{\mp}\pi^{\pm}\pi^{0}$, and $K^{\pm}\pi^{\mp}\pi^{\pm}\pi^{\mp}$ within the invariant mass ranges of [1.845, 1.885] $\rm{GeV}$$/c^{2}$, [1.833, 1.890] $\rm{GeV}$$/c^{2}$, and [1.850, 1.880] $\rm{GeV}$$/c^{2}$, respectively. 
The $D_{s}^{\pm}$ mesons are reconstructed in the channel $K^{\pm}K^{\mp}\pi^{\mp}$ with an invariant mass requirement of [1.950, 1.990] $\rm{GeV}$$/c^{2}$. 
These ranges are approximately $3\sigma$ on either side of
the nominal mass. The $D$ and $D_{s}^{\pm}$ momenta are recalculated from a fit of the momenta of its decay products that constrains them to a common origin and their mass to the nominal mass of the $D$ and $D_{s}^{\pm}$, respectively. 
The $B$ mesons candidates are selected to have a beam-energy-constrained mass larger than $5.27~\rm{GeV}$/$c^{2}$ and energy difference in the range of [$-0.4$, 0.4] $\rm GeV$.

To suppress the background originating from $e^{+}e^{-} \rightarrow q\Bar{q}$ events, a boosted decision
tree~(BDT) classifier is used to veto candidates from continuum events,
training them on equal numbers of simulated signal and continuum.
Differing from the invisible search, for $a' \rightarrow \gamma\gamma$, the nominal approach is to scan the resonances directly on the invariant mass spectrum of the two photons. 

The signal PDFs are determined by using simulated signal samples with different mass hypotheses. Generally, a combination of two gaussians and one asymmetric gaussian is fitted to describe the signal shape, which is well-defined. The background is described by a second order polynomial function.

\begin{figure}[thb]
    \centering
    \includegraphics[width=10cm]{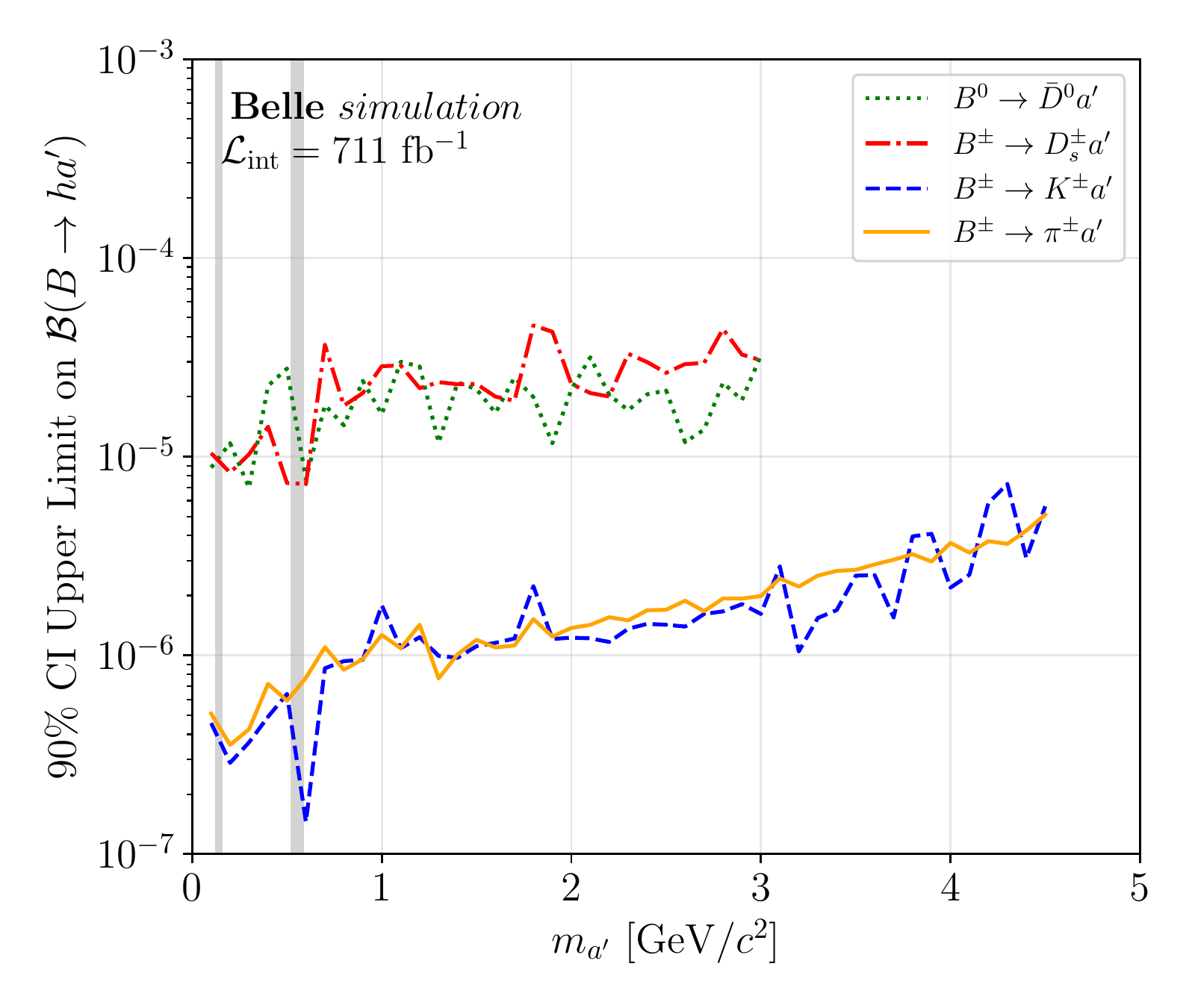}
    \caption{Expected upper limit for $B \rightarrow h (a' \rightarrow \gamma \gamma )$ at Belle experiment, where $h =$ $\pi,~K,~D^{0},~D_{s}$. The filled grep bands are the veto ranges of $\pi^{0} \to 2\gamma$ and $\eta \to 2\gamma$, respectively}
    \label{fig:sens_plot_diphoton}
\end{figure}

An unbinned maximum likelihood fit is performed on the background-only simulated sample. During the fit, the background yield is the only parameter allowed to float and a $90\%$ confidence interval is calculated which are used to estimate the branching fraction upper limit. Fig~\ref{fig:sens_plot_diphoton} shows the corresponding branching fraction upper limit result from Belle simulation. 


%% file: Belle-BelleII.tex
\begin{figure}[!t]
  \begin{center}
    
    \begin{overpic}[width=0.46\textwidth]{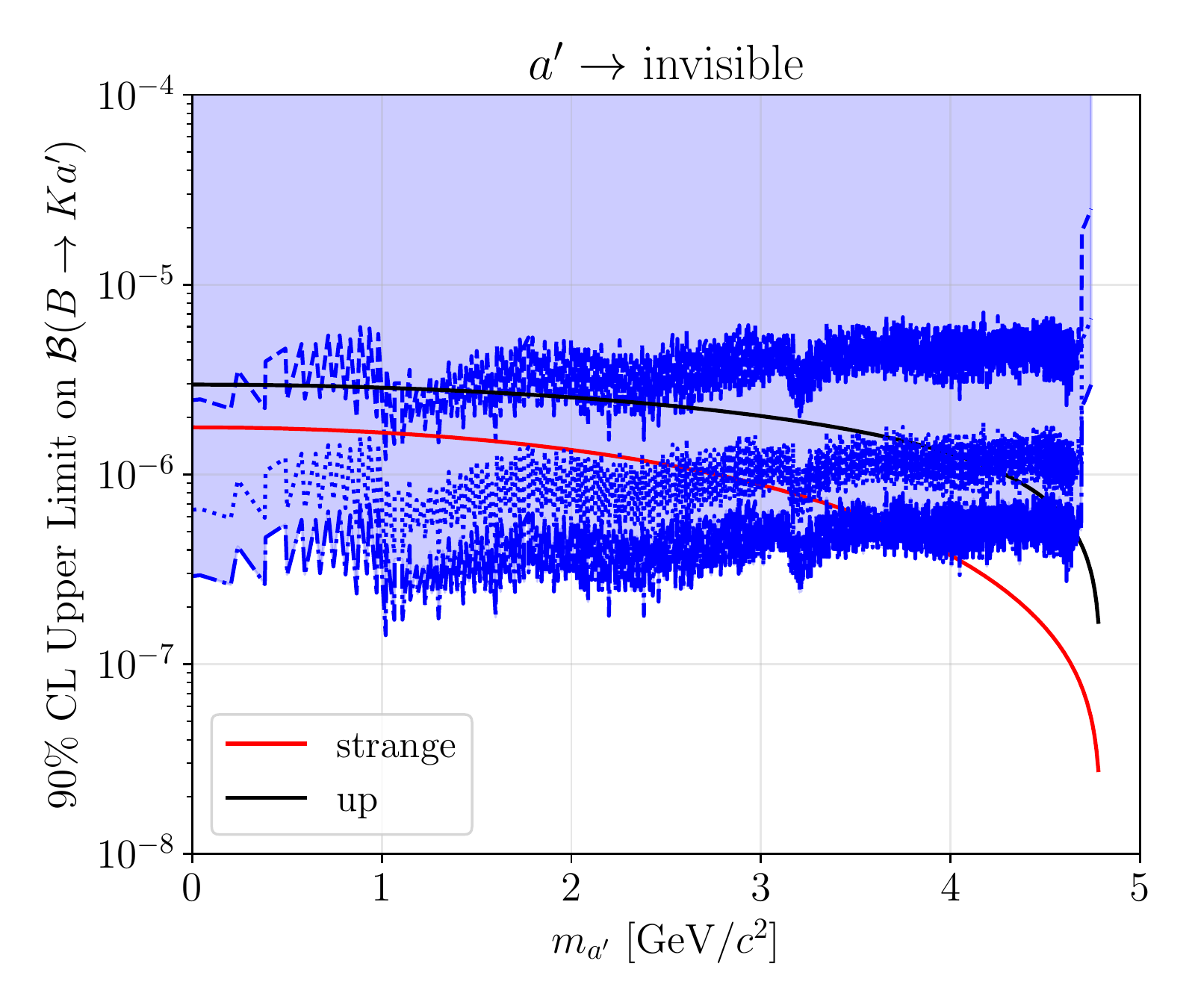}  \put(20,70){(a)} \end{overpic} 
    \begin{overpic}[width=0.46\textwidth]{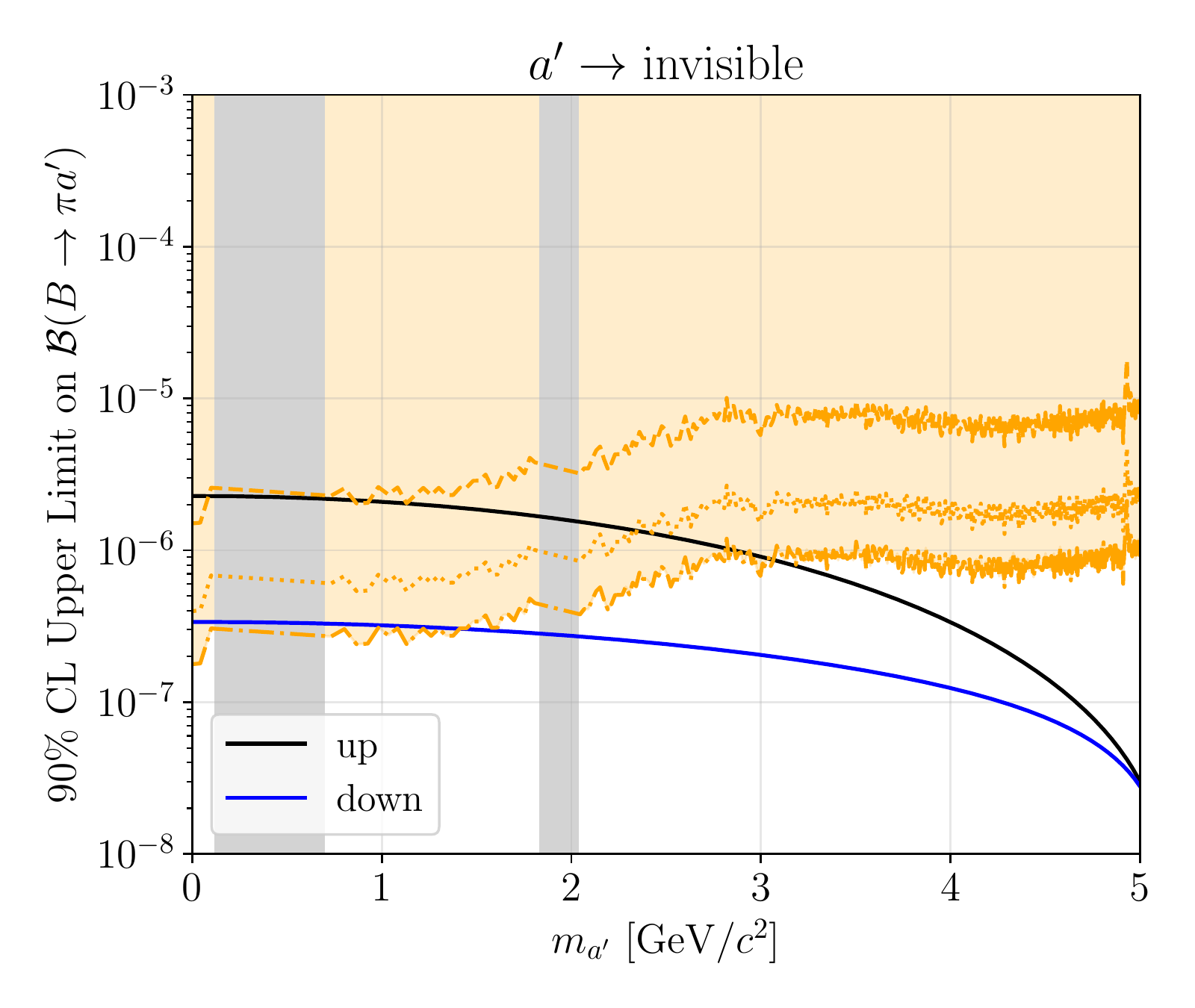}  \put(20,70){(b)} \end{overpic}

    \begin{overpic}[width=0.46\textwidth]{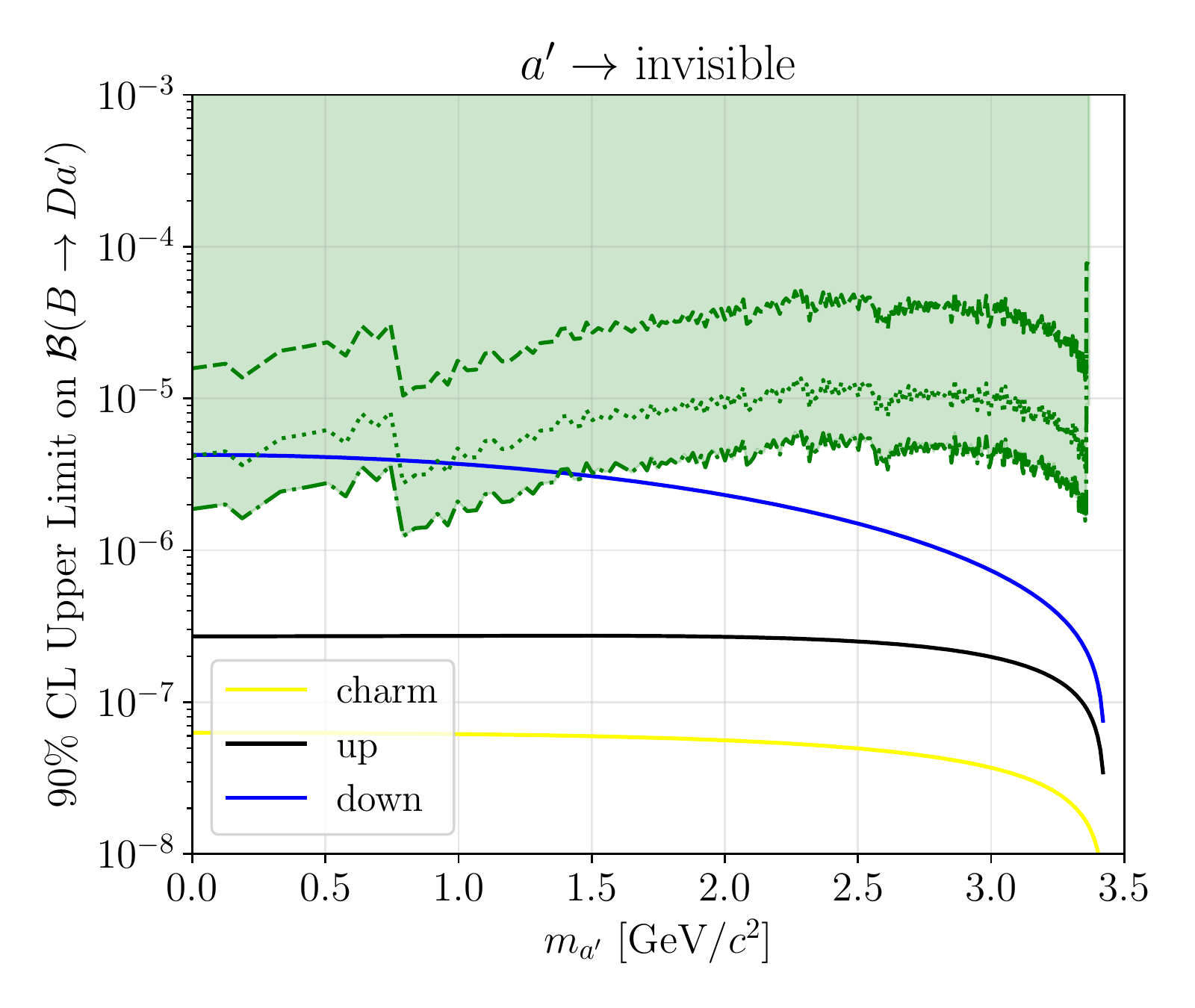}  \put(20,70){(c)} \end{overpic}
    \begin{overpic}[width=0.46\textwidth]{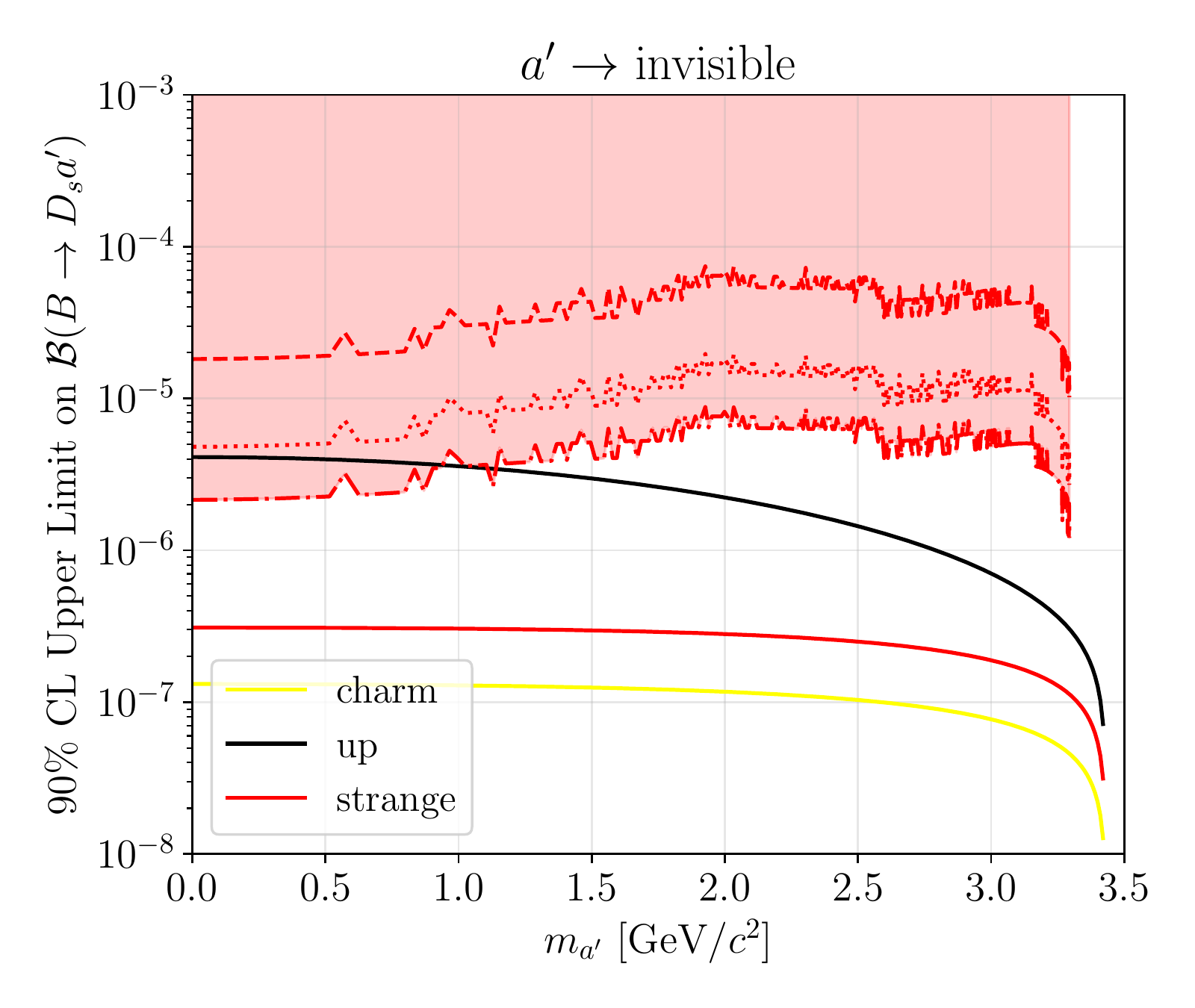}  \put(20,70){(d)} \end{overpic}

  \end{center}
  \caption{Belle and Belle II sensitivity compared with the theoretical prediction for $B \rightarrow h(a^{'} \rightarrow \mathrm{invisible}) $. The exclusion regions in blue, yellow, green and red, are those for $B\to ha'$ with $h=K^\pm, \pi^\pm, D^0, D_s$, respectively. The dash line is simulated Belle sensitivity; The dotted and dash dotted lines are the extrapolation of $\rm 5~ab^{-1}$ and $\rm 10~ab^{-1}$, respectively. The coloured solid lines are theory predictions with $g_i=1$  ($i=u, d, s, c$). The filled grep bands are the veto ranges of $B \to K^{0} \pi$ and $B \to D^{(*)0} \pi$, respectively.}

  \label{fig:invisible2}
\end{figure}
\begin{figure}[!t]
  \begin{center}
    
    \begin{overpic}[width=0.46\textwidth]{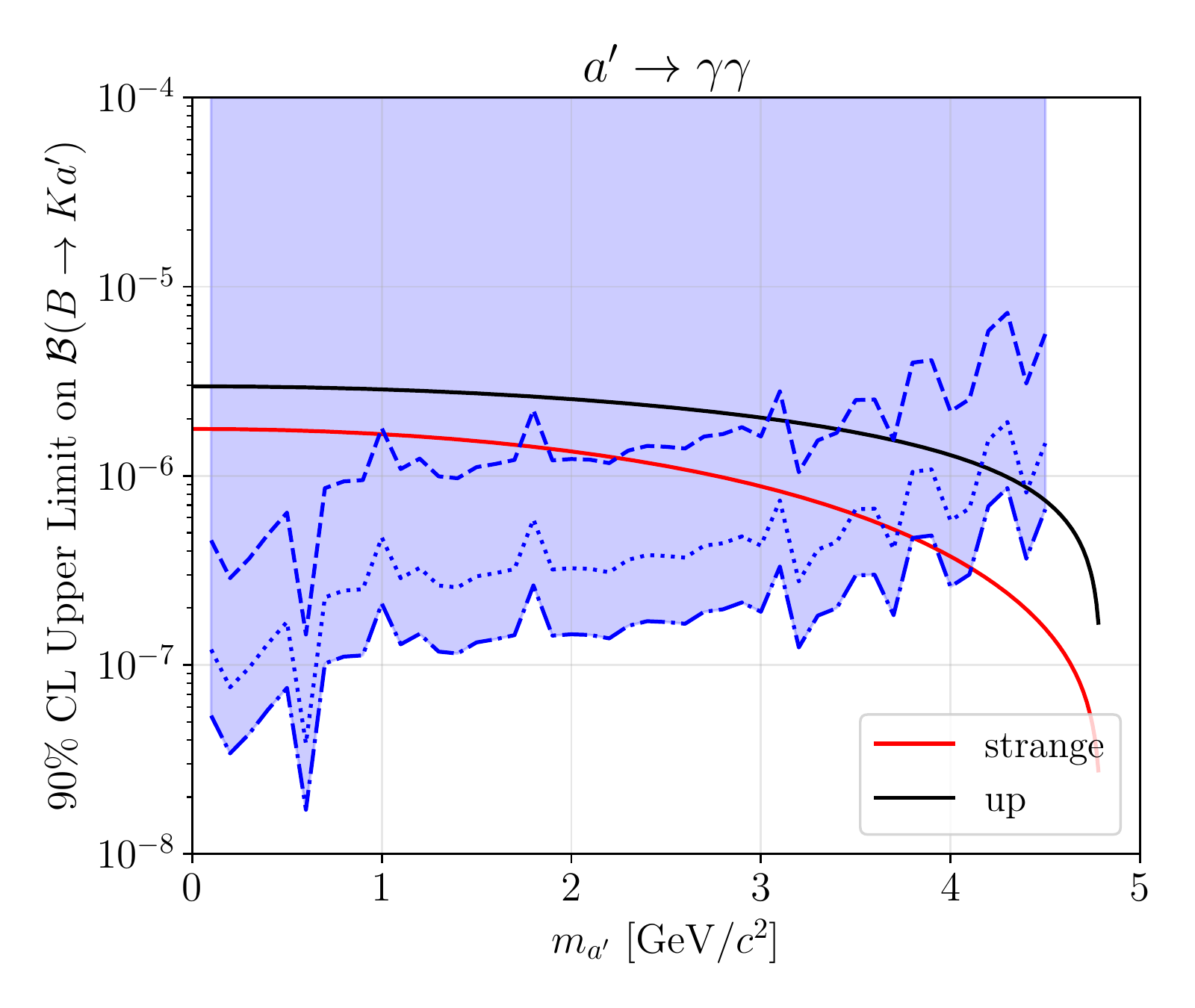}  \put(20,70){(a)} \end{overpic} 
    \begin{overpic}[width=0.46\textwidth]{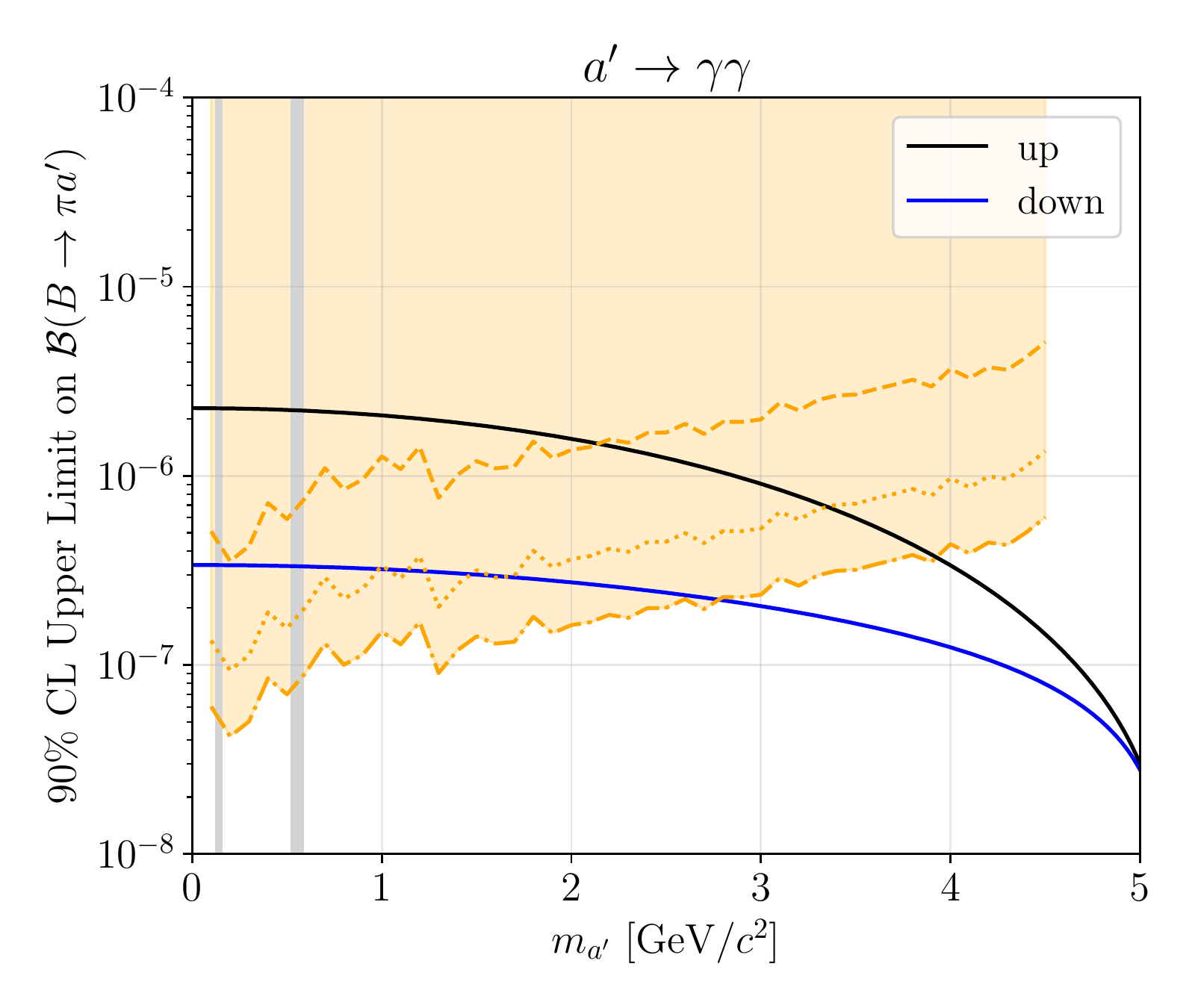}  \put(20,70){(b)} \end{overpic}

    \begin{overpic}[width=0.46\textwidth]{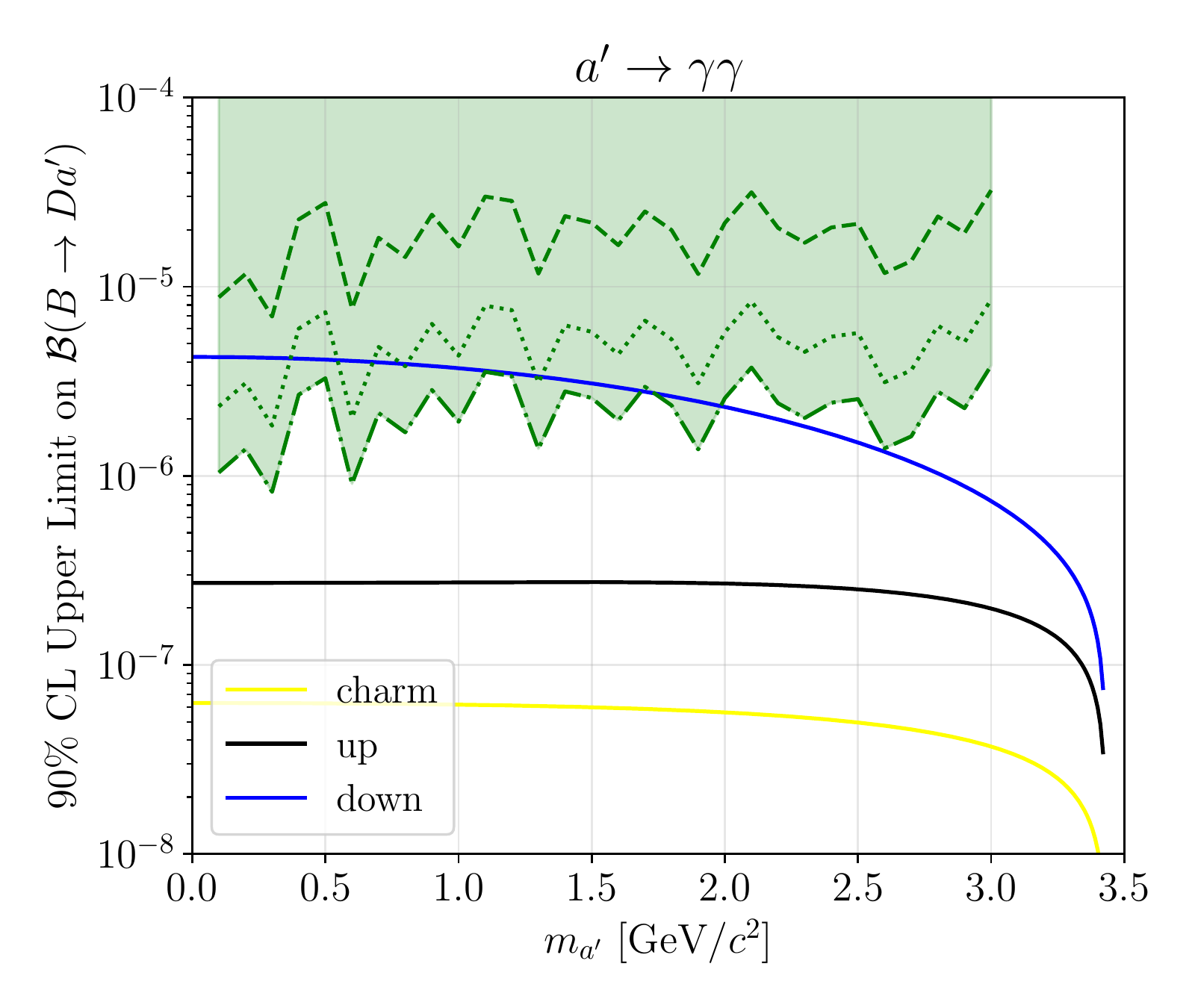}  \put(20,70){(c)} \end{overpic}
    \begin{overpic}[width=0.46\textwidth]{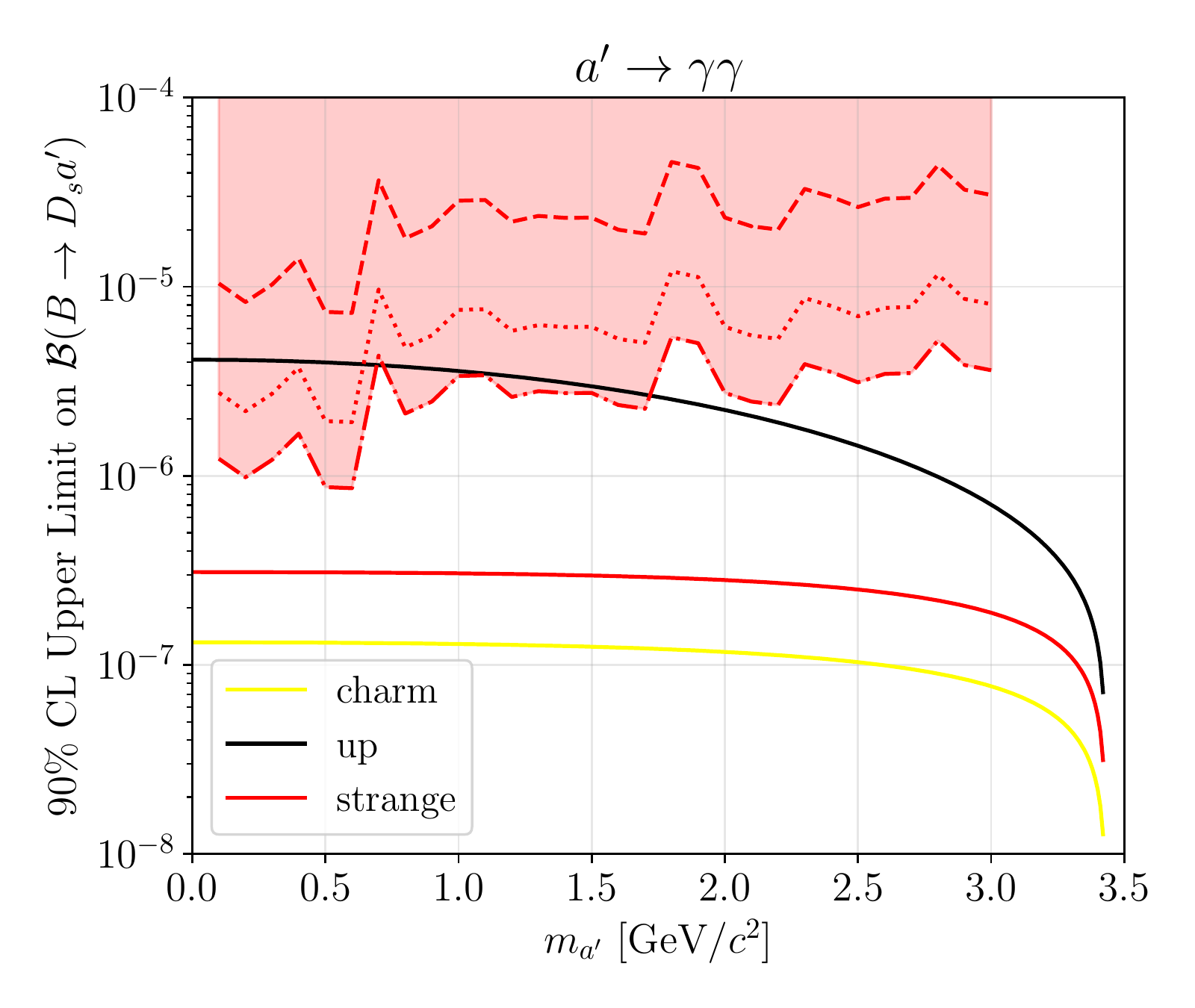}  \put(20,70){(d)} \end{overpic}

  \end{center}

  \caption{Belle and Belle II sensitivity compares with the theoretical prediction for $a^{'} \rightarrow \gamma \gamma$. The exclusion regions in blue, yellow, green and red, are those for $B\to ha'$ with $h=K^\pm, \pi^\pm, D^0, D_s$, respectively. The dash line is simulated Belle sensitivity; The dotted and dash dotted lines are the extrapolation of $\rm 5~ab^{-1}$ and $\rm 10~ab^{-1}$, respectively. The coloured solid lines are theory predictions with $g_i=1$  ($i=u, d, s, c$). The filled grep bands in (b) are the veto ranges of $\pi^{0} \to 2\gamma$ and $\eta \to 2\gamma$, respectively.}

  \label{fig:diphoto2}
\end{figure}

In this section, we discuss the observation potential of ALP produced from $B$ meson annihilation decays at Belle and Belle II. 
In Figs~\ref{fig:invisible2} and~\ref{fig:diphoto2}, we overlay the achievable upper limit of the branching ratios for each channel at Belle and Belle II (as obtained in the previous subsections) and pQCD predictions of the branching ratios (as obtained in Section~\ref{section:2}). 
The Belle II upper limits show in the Figs~\ref{fig:invisible2} and~\ref{fig:diphoto2} are the extrapolation results from Belle.  

Fig~\ref{fig:invisible2} shows that for the invisible final state with the Belle dataset, the ALP that couples to up  quarks with $g_{u}=1$ are in the observable range for $B\to K^\pm a'$ and $B\to \pi^\pm a'$ channels. The strange type model, $g_{s}=1$, can also be accessed with the $B\to K^\pm a'$  channel with Belle II in a few years time (i.e. 10 ab$^{-1}$). The models with non-zero down type coupling, $g_d=1$, might be accessible with the full Belle II dataset via the $B\to D^{0}a'$ channel. 
The $B\to D^{0}a'$ and $B\to D_sa'$ channels are sensitive to the models where ALP couples to charm quark, though $g_c=1$ is too small to be observed at Belle (II) unless there are some new physics model that allow $g_c$ to be enlarged.

From Fig~\ref{fig:diphoto2}, a similar conclusion can be drawn for the di-photon final state: the ALP that couples to up or strange quarks with $g_{u, s}=1$ can already be observed in $B\to K^\pm a'$ channel at Belle and the down type model, $g_{d}=1$, especially with lower mass of ALP, may be accessible in the $B\to D^{0}a'$ or $B\to \pi^\pm a'$ channels with Belle II. For the models where ALP couples to a charm quark, the best channel is $B\to D_s a'$ decay but again, a model that allows a large $g_c$ is needed. 

%% file: pQCD.tex
\section*{Appendix A: Form factor computations in pQCD}
In this section, we briefly describe the computation of the form factors, $f_{1-7}$ in pQCD. We first define the kinematics in the light-cone frame\footnote{The light-cone frame is defined by $p\equiv(p_+,p_-,p_\perp)=(\frac{p_0+p_3}{\sqrt{2}},\frac{p_0-p_3}{\sqrt{2}},(p_1,p_2))$ . The product is given as  $p\cdot p^{\prime}=p_+p^{\prime}_-+p_-p^{\prime}_+-p_\perp\cdot p^{\prime}_\perp$}. First, we define the momentum of the initial and the final states. Here we ignore the term $m_{D^{(*)}}/m_B^2$ and $m_{a'}/m_B^2$: 
\begin{equation}
p_B=\frac{m_B}{\sqrt{2}}(1,1,0), \quad p_{a'}=\frac{m_B}{\sqrt{2}}(0,1,0), \quad p_{D^{(*)}}=\frac{m_B}{\sqrt{2}}(1,0,0) 
\end{equation}
Next, the momentum of the spectator quark of $B$ meson in Fig. 1(a), which we call it as $k_1$ and the spectator of $D^{(*)}$ meson in Fig. 1(b), which we call as $k_2$
\begin{equation}
k_1=(\frac{m_B}{\sqrt{2}}x_1,0,k_\perp^1), \quad k_2=(\frac{m_B}{\sqrt{2}}x_2,0,k_\perp^2) 
\end{equation}
The longitudinal and transverse polarisations of the $D^*$ can be given, respectively, as
\begin{equation}
\epsilon^L=\frac{1}{\sqrt{2}}\frac{m_B}{m_{D^*}}(1,0,0), \quad \epsilon^T=(0,0,\epsilon_\perp) 
\end{equation}

We use the following spin projection and the wave function to describe the $B$ and $D^{(*)}$ mesons: 
\bea
\int\frac{d^4w}{(2\pi)^4}e^{ik \omega}\langle 0 | \bar{d}_{\alpha}(0)b_\delta (w)|\Bbar \rangle
&=&\frac{-i}{\sqrt{2N_C}}\Big[(\Slash{p}_B+m_B)\gamma_5\Big]_{\delta\alpha}\phi_B(k) \\
\int\frac{d^4w}{(2\pi)^4}e^{ik \omega}\langle D^0 | \bar{c}_{\alpha}(0)u_\delta (w)|0 \rangle
&=&\frac{i}{\sqrt{2N_C}}\Big[\gamma_5(\Slash{p}_D+m_D)\Big]_{\delta\alpha}\phi_D(k) \\
\int\frac{d^4w}{(2\pi)^4}e^{ik \omega}\langle D^{0*} | \bar{c}_{\alpha}(0)u_\delta (w)|0 \rangle
&=&\frac{1}{\sqrt{2N_C}}\left\{\Big[\Slash{\epsilon}_{D^*}^*\Slash{p}_{D^*}\Big]_{\delta\alpha}\phi_{D^*_T}(k) +\Big[\Slash{\epsilon}_{D^*}^*m_{D^*}\Big]_{\delta\alpha}\phi_{D^*_L}(k)\right\}\nonumber \\
\int\frac{d^4w}{(2\pi)^4}e^{ik \omega}\langle \Km | \bar{s}_{\alpha}(0)u_\delta (w)|0 \rangle
&=&\frac{i}{\sqrt{2N_C}}\Big[\gamma_5\Slash{p}_K\phi_K(k)+\gamma_5\mu_0\phi_K^{p}(k)+\gamma_5(\Slash{n}_+\Slash{n}_--1)\mu_0\phi_K^{t}(k)\Big]_{\delta\alpha} \nonumber \\
\int\frac{d^4w}{(2\pi)^4}e^{ik \omega}\langle \Kstbar | \bar{s}_{\alpha}(0)u_\delta (w)|0 \rangle
&=&\frac{1}{\sqrt{2N_C}}\Big[\Slash{\epsilon}_{K^*}\Slash{p}_{K^*}\phi_{K^*}^t(k)+\Slash{\epsilon}_{K^*}m_{K^*}\phi_{K^*}(k)+m_{K^*}\phi_{K^*}^{s}(k)\Big]_{\delta\alpha} \nonumber \\
\eea
where $v$ is the unit vector in direction of $p_K$ and $n_-$ is the opposite direction. 
The meson distribution amplitudes are  
\bea
\phi_B(k)&=&N_B x^2 (1 - x)^2 exp\left[-\frac{1}{2} \left(\frac{xm_B}{\omega_B}\right)^2 -  \frac{\omega_B^2b^2}{2}\right] \\
\phi_{D^{(*)}}(k)&=&\frac{3}{\sqrt{2 N_c}} f_{D^{(*)}} x (1 - x) (1 + C_{D^{(*)}} (1 - 2 x))  \\
\phi_{K}(k)&=& \frac{3}{\sqrt{2 N_c}} f_{K} x (1 - x)(1 + 3 a_1^K (1 - 2 x) + \frac{3}{2} a_2^K (5 (1 - 2 x)^2 - 1)) \\
\phi_{K}^p (k)&=& \frac{1}{2\sqrt{2 N_c}} f_{K} (1 + (30 \eta_3 - \frac{5}{2} \rho^2_K)\frac{1}{2} (3 (1 - 2 x)^2 - 1)  \\
&&+ (-3 \eta_3 \omega_3 - \frac{27}{20} \rho^2_K - \frac{81}{10} \rho^2_K a_2^K)\frac{1}{8} (35 (1 - 2 x)^4 - 30 (1 - 2 x)^2 + 3)) \nonumber\\
\phi_{K}^t (k)&=& \frac{3}{\sqrt{2 N_c}} f_{K} (1 - 2 x) (\frac{1}{6} + (5 \eta_3 - \frac{1}{2} \eta_3 \omega_3 - \frac{7}{20} \rho^2_K - \frac{3}{5} \rho^2_K a_2^K) (1-10x+10 x^2))
\nonumber   \\
&&\\
\phi_{K^*}(k)&=&\frac{3}{\sqrt{2 N_c}} f_{K^*} x(1-x) (1 + 3a_1^{\parallel}(1-2x) + \frac{3}{2}a_2^{\parallel}(5(1-2x)^2-1) \\
\phi_{K^*}^t(k)&=&\frac{1}{2\sqrt{2 N_c}} f_{K^*}^T (3(1-2x)^2+\frac{3}{2}a_1^\perp (1-2x)(3(1-2x)^2-1) \\
&&+\frac{3}{2}a_2^\perp (1-2x)^2(5(1-2x)^2-3)+\frac{35}{4}\zeta_3^T(3-30(1-2x)^2+35(1-2x)^4)) \nonumber\\
\phi_{K^*}^s(k)&=&\frac{3}{\sqrt{2 N_c}} f_{K^*}^T  ( a_1^\perp (1 - 6 x + 6 x^2) + (1 - 2 x) (1 + (a_2^\perp + \frac{70}{3} \zeta_3^T) (1 - 10 x + 10 x^2) )) \nonumber \\
&&\\
\phi_{\pi}(k)&=& \frac{3}{\sqrt{2 N_c}} f_{\pi} x (1 - x)(1 + 0.44\frac{3}{2} (5 (1 - 2 x)^2 - 1)+ 0.25\frac{15}{8} (21(2x-1)^4-14(2x-1)^2+1) \\
\phi_{\pi}^p (k)&=& \frac{1}{2\sqrt{2 N_c}} f_{\pi} (1 + 0.43\frac{1}{2}(3(2x-1)^2-1)+0.09\frac{1}{8}(35(2x-1)^4-30(2x-1)^2+3)\\
\phi_{\pi}^t (k)&=& \frac{1}{2\sqrt{2 N_c}} f_{\pi} (1 - 2 x) (1+0.55(1-10x+10 x^2))   
\eea
with $N_B=92$ GeV, $\omega_B=0.4$ GeV, $C_{D^{(*)}}=0.7$. This wave function is normalised to match to the definitions of the  decay constants  and their derivatives introduced earlier: 
\bea
\langle {0}|  \overline{d}\gamma_\mu \gamma_5 b| {\Bbar} \rangle &\equiv&i f_B p^{\mu}_B \\ 
\langle \Dbar| \overline{c}\gamma_\mu \gamma_5u | 0 \rangle &\equiv& -i f_D p^{\mu}_D \\
\langle \Dstbar_L|  \overline{c}\gamma_\mu u | 0 \rangle &\equiv& f_{D^*} m_{D^*} \epsilon^{*}_{\mu D^*}\\
\langle \Dstbar_T|  \overline{c}\sigma_{\mu\nu} u | 0 \rangle &\equiv& -i f_{D^*}^T ( \epsilon^{*}_{\mu D^*}k_\nu-\epsilon^{*}_{\nu D^*}k_\mu)\\
\langle \Km| \overline{s}\gamma_\mu \gamma_5u | 0 \rangle &\equiv& -i f_K p^{\mu}_K \\
\langle \Km| \overline{s}\gamma_5u) | 0 \rangle &\equiv& i \mu_0 f_K  
\eea

The distribution function, $\Phi_{B,D^{(*)}}(t)=\phi_{B,D^{(*)}}(x,b)S_t(x)exp[-S_B(x, b, t)]$ includes the threshold factor, $S_t(x)$ and the Sukakov form factor, $exp[-S_{B,D^{(*)}}(x,b,t)]$ which are 
\bea
S_t(x)&=&\frac{2^{(1+2c)}\Gamma(3/2+c)}{\sqrt{\pi}\Gamma(1+c)} (x(1-x))^c\\
S_B(x_1, b_1, t)&=&\left\{\begin{array}{ccc}s(\frac{x_1m_B}{\sqrt{2}},b_1)+2\int^{t}_{1/b_1}\frac{d\mu}{\mu} \ \gamma(\mu)
&  \ & {\rm for}\ S_B  >0
\\
0 &  \ & {\rm for}\ S_B< 0
\end{array}\right.\\
S_{D^{(*)}}(x_2, b_2, t)&=&\left\{\begin{array}{ccc}s(\frac{x_2m_B}{\sqrt{2}},b_2)+2\int^{t}_{1/b_2}\frac{d\mu}{\mu} \ \gamma(\mu)
&  \ & {\rm for}\ S_{D^{(*)}}  >0
\\
0 &  \ & {\rm for}\ S_{D^{(*)}}< 0
\end{array}\right.\\
S_{K}(x_2, b_2, t)&=&\left\{\begin{array}{ccc}s(\frac{x_2m_B}{\sqrt{2}},b_2)+s(\frac{(1-x_2)m_B}{\sqrt{2}},b_2)+2\int^{t}_{1/b_2}\frac{d\mu}{\mu} \ \gamma(\mu)
&  \ & {\rm for}\ S_K  >0
\\
0 &  \ & {\rm for}\ S_K< 0
\end{array}\right.\\
s(Q,b)&=&\left\{\begin{array}{ccc}
\int_{1/b}^{Q}\frac{d\mu}{\mu} \ \ln[\frac{Q}{\mu}]A(\mu)+B(\mu) & \ & {\rm for}\ Q > 1/b\\
0 &  \ & {\rm for}\ Q < 1/b
\end{array}\right.
\eea
where $c=0.4$ and $\gamma(\mu)=-\frac{\alpha_s(\mu)}{\pi}$.

The $V-A$ form factor for $Ka'$ and $S\pm P$ form factors for $Ka'$ can be obtained in a similar manner\footnote{Note it turned out that the tensor term is small numerically. Thus, we do not write the result explicitly in the following.}. 
\bea
 \langle \Km a' |  (\overline{s}_iu_i)_{V-A} | 0 \rangle
 &=& \int_0^1 dx \int_0^\infty d|b| (ig_u)N_c\left(\frac{i}{\sqrt{2N_c}}\phi_K^p(k)\right)\times (-1)\nonumber \\
&&4\mu_0{\Big[-x_2p_K^\mu-p_{a'}^\mu\Big]}|b_2|\frac{i\pi}{2}H_0(m_B\sqrt{x_2}, |b_2|)\\
\langle a' | (\overline{d}_jb_j)_{S-P} | \Bbar \rangle  
&=& \int_0^1 dx \int_0^\infty d|b| (ig_u)N_c\left(\frac{-i}{\sqrt{2N_c}}\phi_B(k)\right) \times (-1) \nonumber \\
&&4\Big[-x_1(p_B\cdot p_K)+(p_B\cdot p_{a'})\Big]|b_1|K_0(m_B\sqrt{x_1}, |b_1|)\\
\langle \Km a' |  (\overline{s}_iu_i)_{S+P} | 0 \rangle
 &=& \int_0^1 dx \int_0^\infty d|b| (ig_u)N_c\left(\frac{i}{\sqrt{2N_c}}\phi_K(k)\right)\times (-1)\nonumber \\
&&4\mu_0{\Big[x_2m_K^2+(p_K\cdot p_{a'})\Big]}|b_2|\frac{i\pi}{2}H_0(m_B\sqrt{x_2}, |b_2|)
\eea

The integration must be taken in the range $\frac{\sqrt{2}}{m_B}\Lambda_{QCD}<x<1, 0<b<\frac{1}{\Lambda_{QCD}}$ after multiplying with the Wilson coefficient, 
\be
a_1(t)\quad {\rm with}\quad   t=max[\sqrt{x}m_B,1/b]
\ee

\section*{Appendix B: Computation of the penguin process $\Bm \to \Km a'$ in pQCD}
The Weak Hamiltonian which gives the $\Bm \to \Km a'$  process is 
\be
H^{\rm eff}=\frac{G_F}{\sqrt{2}}\Big[V_{ub}^*V_{us}(C_1 O_1+C_2 O_2) - V_{tb}^*V_{ts}\sum_{i=1,10}C_i O_i\Big]
\ee
where the operators $O_i$ are defines as 
\bea
O_1=(\overline{s}_iu_j)_{V-A}(\overline{u}_jb_i)_{V-A}, &\quad& O_2=(\overline{s}_iu_i)_{V-A}(\overline{u}_jb_j)_{V-A} \nonumber \\
O_{3}=(\overline{s}_ib_i)_{V-A}(\overline{u}_ju_j)_{V-A}, &\quad& O_{4}=(\overline{s}_ib_j)_{V-A}(\overline{u}_ju_i)_{V-A}  \nonumber \\
O_{5}=(\overline{s}_ib_i)_{V-A}(\overline{u}_ju_j)_{V+A}, &\quad& O_{6}=(\overline{s}_ib_j)_{V-A}(\overline{u}_ju_i)_{V+A}  \nonumber \\
O_{7}=\frac{3}{2}e_u(\overline{s}_ib_i)_{V-A}(\overline{u}_ju_j)_{V+A}, &\quad& O_{8}=\frac{3}{2}e_u(\overline{s}_ib_j)_{V-A}(\overline{u}_ju_i)_{V+A}  \nonumber \\
O_{9}=\frac{3}{2}e_u(\overline{s}_ib_i)_{V-A}(\overline{u}_ju_j)_{V-A}, &\quad& O_{10}=\frac{3}{2}e_u(\overline{s}_ib_j)_{V-A}(\overline{u}_ju_i)_{V-A}  
\eea
In order to have the initial $b$ quark and $\overline{u}$ quark be in the same current (see Fig.\ref{fig:1}), $O_{3,10}$ must be  Fiertz transformed in the Dirac space. 
\bea
O^F_{3, 9}\propto (\overline{u}_jb_i)_{V-A}(\overline{s}_iu_j)_{V-A}, &\quad& O^F_{4, 10}\propto (\overline{u}_jb_j)_{V-A}(\overline{s}_iu_i)_{V-A}  \nonumber \\
O^F_{5, 7}\propto -2(\overline{u}_jb_i)_{S-P}(\overline{s}_iu_j)_{S+P}, &\quad& O^F_{6, 8}\propto -2(\overline{u}_jb_j)_{S-P}(\overline{s}_iu_i)_{S+P}  
\eea
where $(\overline{\psi}\psi^\prime)_{S\pm P}=\overline{\psi}(1\pm\gamma_5)\psi^\prime$
After applying the Fiertz transformation to the colour space, we find the combination of the Wilson coefficients $a_2O_2$, $a_4O_4^F$ and $a_6O_6^F$ where $a_4=C_4+C_3/N_C+C_{10}+C_{9}/N_C, a_6=C_6+C_5/N_C+C_8+C_7/N_C$.

The amplitude of $\Bm \to \Km a'$ can be obtained by sandwiching this Hamiltonian by the initial and the final states: 
\be
A=\frac{G_F}{\sqrt{2}}\Big\{ V_{ub}^*V_{us}{a_2}\langle \Km a' | O_2 | \Bm \rangle
- V_{tb}^*V_{ts}\Big[{a_4}\langle \Km a' | O_4^F | \Bm \rangle+{a_6}\langle \Km a' | O_6^F | \Bm \rangle\Big] \Big\} 
\ee
The amplitudes for the APLs emission from the initial $b$ and $\overline{u}$ and the final $s$ and $\overline{u}$ in Fig.~1 are given, respectively
\bea
A^{1a}&=&\frac{G_F}{\sqrt{2}}\Big\{V_{ub}^*V_{us}{a_2}\langle \Km| (\overline{s}_iu_i)_{V-A} | 0 \rangle \langle a' | (\overline{u}_jb_j)_{V-A} | \Bm \rangle  \\
&&\hspace*{1cm}-V_{tb}^*V_{ts}{a_{4}}\langle \Km| (\overline{s}_iu_i)_{V-A} | 0 \rangle \langle a' | (\overline{u}_jb_j)_{V-A} | \Bm \rangle \nonumber \\
&&\hspace*{1cm}-V_{tb}^*V_{ts}(-2a_6)\langle \Km| (\overline{s}_iu_i)_{S+P} | 0 \rangle \langle a' | (\overline{u}_jb_j)_{S-P} | \Bm \rangle   \Big\} \\ 
A^{1b}&=&\frac{G_F}{\sqrt{2}}\Big\{V_{ub}^*V_{us}{a_2} \langle \Km a' | (\overline{s}_iu_i)_{V-A} | 0 \rangle \langle 0 | (\overline{u}_jb_j)_{V-A} | \Bm \rangle \\ 
&&\hspace*{1cm}-V_{tb}^*V_{ts}{a_{4}}\langle \Km a' | (\overline{s}_iu_i)_{V-A} | 0 \rangle \langle 0 | (\overline{u}_jb_j)_{V-A} | \Bm \rangle  \nonumber \\
&&\hspace*{1cm}-V_{tb}^*V_{ts}(-2a_6)\langle \Km a' | (\overline{s}_iu_i)_{S+P} | 0 \rangle \langle 0 | (\overline{u}_jb_j)_{S-P} | \Bm \rangle  \Big\} 
\eea
Then, after defining the form factors as\footnote{Note that the following expressions are correct up to the tensor wave function contribution. The tensor term makes the expression very messy while its contribution turned out to be very small numerically. Thus, we refrain from writing down the full expression.}
\bea
\langle a' | (\overline{u}_jb_j)_{V-A} | \Bm \rangle  &\equiv& p^{\mu}_K f_1^{\Km}+p^{\mu}_{a'} f_2^{\Km}\\
\langle \Km a' |  (\overline{s}_iu_i)_{V-A} | 0 \rangle  &\equiv& p^{\mu}_K f_3^{\Km}+p^{\mu}_{a'} f_4^{\Km} \\
\langle a' | (\overline{u}_jb_j)_{S-P} | \Bm \rangle  &\equiv& (p_K\cdot p_B) f_5^{\Km}+ (p_{a'}\cdot p_B)f_6^{\Km}\\
\langle \Km a' |  (\overline{s}_iu_i)_{S+P} | 0 \rangle  &\equiv&(p_K\cdot p_K) f_7^{\Km}+ (p_{a'}\cdot p_K)f_8^{\Km} 
 \eea   
We can obtain the amplitudes as: 
 \bea
 A^{1a}(\Bm\to \Km a') &=& \frac{G_F}{\sqrt{2}}\Big\{ V_{ub}^*V_{us}(i f_K) {a_2} \left[f_1^{\Km} (p_K\cdot p_K)+f_2^{\Km} (p_K\cdot p_{a'} )\right] \\
 &&\hspace*{1cm} -V_{tb}^*V_{ts}(i f_K) {a_4} \left[f_1^{\Km} (p_K\cdot p_K)+f_2^{\Km} (p_K\cdot p_{a'} )\right] \nonumber \\
 &&\hspace*{1cm} -  V_{tb}^*V_{ts}\left(i \mu_0  f_K  \right)(-2a_6)  \left[f_5^{\Km}(p_K\cdot p_B) + f_6^{\Km}(p_{a'}\cdot p_B)\right]\Big\} \nonumber\\
  A^{1b}(\Bm\to \Km a') &=& \frac{G_F}{\sqrt{2}}\Big\{ V_{ub}^*V_{us}(-i f_B) {a_2} \left[f_3^{\Km} (p_B\cdot p_K)+f_4^{\Km} (p_B\cdot p_{a'} )\right] \\
 &&\hspace*{1cm} -V_{tb}^*V_{ts}(-i f_B) {a_4} \left[f_3^{\Km} (p_B\cdot p_K)+f_4^{\Km} (p_B\cdot p_{a'} )\right] \nonumber \\
 &&\hspace*{1cm} -  V_{tb}^*V_{ts}\left( im_B f_B \right) (-2a_6) \left[ f_7^{\Km}(p_K\cdot p_K)+ f_8^{\Km}(p_{a'}\cdot p_K) \right]\Big\} \nonumber
 \eea